\documentclass[prd,preprint,superscriptaddress,showpacs,byrevtex]{revtex4}
\usepackage{bm}
\def\fsl#1{\setbox0=\hbox{$#1$}           
   \dimen0=\wd0                                 
   \setbox1=\hbox{/} \dimen1=\wd1               
   \ifdim\dimen0>\dimen1                        
      \rlap{\hbox to \dimen0{\hfil/\hfil}}      
      #1                                        
   \else                                        
      \rlap{\hbox to \dimen1{\hfil$#1$\hfil}}   
      /                                         
   \fi}                                         %
\newcommand{\be}{\begin{equation}}
\newcommand{\ee}{\end{equation}}
\newcommand{\bea}{\begin{eqnarray}}
\newcommand{\eea}{\end{eqnarray}}
\newcommand{\beq}{\begin{equation}}
\newcommand{\eeq}{\end{equation}}
\newcommand{\beqs}{\begin{eqnarray}}
\newcommand{\eeqs}{\end{eqnarray}}

\begin{document}
\title{ Proof of NRQCD Color Octet Factorization of P-Wave Heavy Quarkonium Production In Non-Equilibrium QCD at RHIC and LHC }
\author{Gouranga C Nayak }\thanks{E-Mail: nayakg138@gmail.com}
\date{\today}
\begin{abstract}
Recently we have proved NRQCD color octet factorization of S-wave and P-wave heavy quarkonia production at all orders in coupling constant in QCD in vacuum at high energy colliders in \cite{nksw} and in \cite{nkpw} respectively. In this paper we extend this to prove NRQCD color octet factorization of P-wave heavy quarkonium production in non-equilibrium QCD at RHIC and LHC at all orders in coupling constant. This proof is necessary to study the quark-gluon plasma at RHIC and LHC.
\end{abstract}
\maketitle
\pagestyle{plain}
\pagenumbering{arabic}
\section{Introduction}
Recently we have proved NRQCD color octet factorization of S-wave and P-wave heavy quarkonia production at all orders in coupling constant in QCD in vacuum at high energy colliders in \cite{nksw} and in \cite{nkpw} respectively. In this paper we extend this to prove NRQCD color octet factorization of P-wave heavy quarkonium production in non-equilibrium QCD at RHIC and LHC at all orders in coupling constant. This proof is necessary to study the quark-gluon plasma at RHIC and LHC.

During the early stage of the universe, just after $10^{-12}$ seconds of the big bang, our universe was filled with a hot and dense state of matter known as the quark-gluon plasma (QGP). The temperature of the quark-gluon plasma is $\gtrsim$ 200 MeV ($\gtrsim 10^{12}~^0K$) and it is the densest state of the matter in the universe after the black hole. This temperature $\gtrsim$ 200 MeV of the quark-gluon plasma corresponds to the energy density $\gtrsim$ 2 GeV/$fm^3$ of the quark-gluon plasma. For the comparison, the energy density of the normal nucleus is $\sim$ 0.15 GeV/$fm^3$. Hence it is important to recreate this early universe scenario in the laboratory, {\it i. e.}, it is important to create the quark-gluon plasma in the laboratory.

There are two experiments in the laboratory which study the production of quark-gluon plasma. One is the RHIC (relativistic heavy-ion colliders) and another is the LHC (large hadron colliders). The RHIC collides two gold nuclei (Au-Au collisions) at the center of mass energy per nucleon $\sqrt{s}_{NN}$ = 200 GeV with the total energy of 200 GeV x 197 = 39.4 TeV. Similarly the LHC collides two lead nuclei (Pb-Pb collisions) at the center of mass energy per nucleon $\sqrt{s}_{NN}$ = 5.02 TeV (in the second run) with the total energy of 5.02 TeV x 208 = 1044.16 TeV.

Since this total energy is very high which is deposited over a very small volume just after the nuclear collisions at RHIC and LHC there is no doubt that the required energy density $\gtrsim$ 2 GeV/$fm^3$ to produce the quark-gluon plasma is reached at RHIC and LHC. However, it is not clear if this quark-gluon plasma is thermalized at RHIC and LHC. This is because of the following reason.

Since the center of mass energy per nucleon ($\sqrt{s}_{NN}$ = 200 GeV) at RHIC and the center of mass energy per nucleon ($\sqrt{s}_{NN}$ = 5.02 TeV) at LHC heavy-ion colliders are very high, the two nuclei at RHIC and LHC travel almost at the speed of light. Since the two nuclei before the nuclear collisions travel almost at the speed of light, the longitudinal momenta of the partons inside the nuclei are much larger than the transverse momenta of the partons. Hence the partons produced just after the nuclear collisions at RHIC and LHC have large momentum anisotropy. Since the partons produced just after the nuclear collisions at RHIC and LHC have large momentum anisotropy, the quark-gluon plasma produced just after the nuclear collisions at RHIC and LHC is in non-equilibrium.

For this non-equilibrium quark-gluon plasma to equilibrate to form the thermalized quark-gluon plasma at RHIC and LHC, many more secondary partonic collisions are necessary. If the time of the non-equilibrium stage is large then there can be many more secondary partonic collisions. However, we do not directly experimentally measure the time of the non-equilibrium stage because all we measure experimentally are hadrons and other non-hadronic color singlet observables. Hence we do not know whether the time of the non-equilibrium stage is large or small. All we know is that the hadronization time scale in QCD is very small $\sim 10^{-24}$ seconds.

Since the two nuclei at RHIC and LHC travel almost at the speed of light and since the hadronization time scale in QCD is very small ($\sim 10^{-24}$ seconds) it is unlikely that there are many more secondary partonic collisions to bring the system into equilibrium. Hence the quark-gluon plasma at RHIC and LHC heavy-ion colliders may be in non-equilibrium from the beginning to the end, {\it i. e.}, the quark-gluon plasma at RHIC and LHC may be in non-equilibrium starting from the initial time just after the nuclear collision to the hadronization stage.

Since the quark-gluon plasma at RHIC and LHC heavy-ion colliders may be in non-equilibrium from the beginning to the end it is necessary to do the theoretical calculation at RHIC and LHC in non-equilibrium QCD. The first principle method to study the non-equilibrium QCD is by using the closed-time path (CTP) integral formalism.

Note that we have not directly experimentally observed quarks and gluons and hence we cannot directly detect the quark-gluon plasma. For this reason the indirect signatures are proposed to detect the quark-gluon plasma. The indirect signatures for the quark-gluon plasma detection at RHIC and LHC are, 1) heavy quarkonium production/suppression, 2) jet quenching, 3) dilepton and direct photon production and 4) strangeness enhancement.

The $j/\psi$ suppression is proposed to be a major signature of the quark-gluon plasma detection \cite{msl}. The basic idea behind the $j/\psi$ suppression in the quark-gluon plasma is the Debye screening phenomena. If the Debye screening length at high temperature becomes smaller than the $j/\psi$ radius then the $j/\psi$ is completely suppressed in the quark-gluon plasma. The Debye screening length was calculated in \cite{msl} by using lattice QCD method at finite temperature.

However, if the quark-gluon plasma is in non-equilibrium then there is no notion of temperature as the temperature is not defined in non-equilibrium. Hence the lattice QCD method at finite temperature is not applicable in non-equilibrium. From this point of view the Debye screening length needs to be calculated by using the nonequilibrium-nonperturbative QCD. However, this may not be necessary because from the experimental data of STAR \cite{rh1}, PHENIX \cite{rh2}, ALICE \cite{lh1}, CMS \cite{lh2} and ATLAS \cite{lh3} collaborations we have not found complete suppression of $j/\psi$ at RHIC and LHC heavy-ion colliders.

Hence it is not necessary to study $j/\psi$ suppression via screening in equilibrium QGP or in non-equilibrium QGP. This is because whether it is equilibrium QGP or non-equilibrium QGP the $j/\psi$ will be completely suppressed by the screening mechanism of Matsui and Satz \cite{msl}. However, as mentioned above, we have not found the complete suppression of $j/\psi$ at RHIC and LHC heavy-ion colliders \cite{rh1,rh2,lh1,lh2,lh3}. This implies that it is not necessary to investigate the $j/\psi$ suppression mechanism via screening \cite{msl} at RHIC and LHC heavy ion colliders. Instead, it is necessary to understand the heavy quarkonium production mechanism at RHIC and LHC heavy-ion colliders.

Note that before understanding the heavy quarkonium production mechanism at RHIC and LHC heavy-ion colliders it is necessary to understand the heavy quarkonium production mechanism in the $pp$ collisions at the same center of mass energy. This because in order to quantify the quark-gluon plasma medium effect, it is necessary to compare the heavy quarkonium production cross section in heavy-ion collisions at RHIC and LHC with the heavy quarkonium production cross section in $pp$ collisions at the same center of mass energy per nucleon.

In $pp$ collisions at the high energy colliders the non-relativistic QCD (NRQCD) color octet mechanism \cite{bodx} is successful in explaining the experimental data. For example, the NRQCD color octet mechanism is successful in explaining the experimental data in $p{\bar p}$ collisions at Tevatron \cite{cd1,cd2}, in $pp$ collisions at LHC \cite{cd3,cd4,cd5,cd6} and in $pp$ collisions at RHIC \cite{cd7,cd8}. Since the NRQCD color octet mechanism explains the experimental data in $pp$ collisions at high energy colliders it is necessary to study the heavy quarkonium production at RHIC and LHC heavy ion colliders by using NRQCD color octet mechanism.

In order to study heavy quarkonium production at high energy colliders in NRQCD color octet mechanism it is necessary to prove factorization of heavy quarkonium production in NRQCD color octet mechanism. This is because if one does not prove the factorization of NRQCD heavy quarkonium production in heavy-ion collisions at RHIC and LHC then one will predict the infinite cross section of heavy quarkonium production. The proof of factorization plays an important role to study physical observables at high energy colliders \cite{csg,nks,nka}.

Note that in the original formulation of the NRQCD heavy quarkonium production \cite{bodx} the proof of factorization in NRQCD color octet mechanism was missing. In \cite{nks} we have proved the NRQCD color octet factorization of S-wave heavy quarkonium production at next-to-next leading order (NNLO) in coupling constant by using the diagrammatic method. In \cite{nksw} we have proved the NRQCD color octet factorization of the S-wave heavy quarkonium production at high energy colliders at all orders in coupling constant by using the path integral formulation of QCD. In \cite{nkpw} we have proved the NRQCD color octet factorization of the P-wave heavy quarkonium production at high energy colliders at all orders in coupling constant by using the path integral formulation of QCD. In this paper we extend this to prove the NRQCD color octet factorization of P-wave heavy quarkonium production in non-equilibrium QCD at RHIC and LHC at all orders in coupling constant. This proof is necessary to study the quark-gluon plasma at RHIC and LHC.

The paper is organized as follows. In section II we describe the non-perturbative correlation function of the heavy quark-antiquarks in non-equilibrium QCD by using the closed-time path integral formalism. In section III we describe the infrared divergence and the light-like eikonal line in QCD. In section IV we discuss the infrared divergences in NRQCD color octet P-wave heavy quarkonium production. In section V we prove NRQCD color octet factorization of P-wave heavy quarkonium production in non-equilibrium QCD at all orders in coupling constant. In section VI we derive the definition of the NRQCD color octet non-perturbative matrix element of the P-wave heavy quarkonium production in non-equilibrium QCD. Section VII contains conclusions.

\section{ Non-Perturbative Correlation Function of Heavy Quark-Antiquarks In Non-Equilibrium QCD by Using Closed-Time Path (CTP) Integral Formalism }

In this section we describe the non-perturbative correlation function of heavy quark-antiquarks in non-equilibrium QCD by using the closed-time path integral formalism. We denote the up quark field by $\psi^u_{kr}(x)$, the down quark field by $\psi^d_{kr}(x)$, the strange quark field by $\psi^s_{kr}(x)$ and the heavy quark field by $\Psi_{kr}(x)$ where $k=1,2,3$ is the color index and $r=+,-$ is the closed-time path index. The non-perturbative correlation function of the heavy quark-antiquarks in non-equilibrium QCD of the type $<in|{\bar \Psi}_{kr}(x'){\overline \nabla }_{x'}T^c_{kl} \Psi_{lr}(x') {\bar \Psi}_{js}(x''){\overline \nabla }_{x''}T^c_{jn} \Psi_{ns}(x'')|in>$ is given by \cite{kcn,kcn1}
\bea
&&<in|{\bar \Psi}_{kr}(x'){\overline \nabla }_{x'}T^c_{kl} \Psi_{lr}(x') {\bar \Psi}_{js}(x''){\overline \nabla }_{x''}T^c_{jn} \Psi_{ns}(x'')|in>=\int [d{\bar \psi}^u_+] [d{\psi}^u_+][d{\bar \psi}^u_-] [d{\psi}^u_-][d{\bar \psi}^d_+] [d{\psi}^d_+]\nonumber \\
&&[d{\bar \psi}^d_-] [d{\psi}^d_-][d{\bar \psi}^s_+] [d{\psi}^s_+][d{\bar \psi}^s_-] [d{\psi}^s_-][d{\bar \Psi}_+] [d{\Psi}_+][d{\bar \Psi}_-] [d{\Psi}_-]\nonumber \\
&&\times {\bar \Psi}_{kr}(x'){\overline \nabla }_{x'}T^c_{kl} \Psi_{lr}(x') {\bar \Psi}_{js}(x''){\overline \nabla }_{x''}T^c_{jn} \Psi_{ns}(x'')\times {\rm det}[ \frac{\delta \partial^\sigma Q_{\sigma +}^h}{\delta \omega^e_+}]\times {\rm det}[ \frac{\delta \partial^\sigma Q_{\sigma -}^h}{\delta \omega^e_-}]\nonumber \\
&& \times ~{\rm exp}[i\int d^4x [-\frac{1}{4} F_{\sigma \lambda }^h[Q_+(x)]F^{\sigma \lambda h}[Q_+(x)] +\frac{1}{4} F_{\sigma \lambda }^h[Q_-(x)]F^{\sigma \lambda h}[Q_-(x)]    -\frac{1}{2\gamma} [\partial^\sigma Q_{\sigma +}^h(x)]^2\nonumber \\
&&+\frac{1}{2\gamma} [\partial^\sigma Q_{\sigma -}^h(x)]^2  +{\bar \psi}^u_{k+}(x)[\delta^{kl}(i{\not \partial} -m^u)+gT^h_{kl}{\not Q}_+^h(x)]\psi^u_{l+}(x)\nonumber \\
&& -{\bar \psi}^u_{k-}(x)[\delta^{kl}(i{\not \partial} -m^u)+gT^h_{kl}{\not Q}_-^h(x)]\psi^u_{l-}(x)   +{\bar \psi}^d_{k+}(x)[\delta^{kl}(i{\not \partial} -m^d)+gT^h_{kl}{\not Q}_+^h(x)]\psi^d_{l+}(x)\nonumber \\
&& -{\bar \psi}^d_{k-}(x)[\delta^{kl}(i{\not \partial} -m^d)+gT^h_{kl}{\not Q}_-^h(x)]\psi^d_{l-}(x) +{\bar \psi}^s_{k+}(x)[\delta^{kl}(i{\not \partial} -m^s)+gT^h_{kl}{\not Q}_+^h(x)]\psi^s_{l+}(x)\nonumber \\
&& -{\bar \psi}^s_{k-}(x)[\delta^{kl}(i{\not \partial} -m^s)+gT^h_{kl}{\not Q}_-^h(x)]\psi^s_{l-}(x)  +{\bar \Psi}_{k+}(x)[\delta^{kl}(i{\not \partial} -M)+gT^h_{kl}{\not Q}_+^h(x)]\Psi_{l+}(x)\nonumber \\
&& -{\bar \Psi}_{k-}(x)[\delta^{kl}(i{\not \partial} -M)+gT^h_{kl}{\not Q}_-^h(x)]\Psi_{l-}(x)  ]]\nonumber \\
&&\times <0,Q^+,\Psi^+,{\bar \Psi}^+,\psi^+_s,{\bar \psi}^+_s,\psi^+_d,{\bar \psi}^+_d,\psi^+_u,{\bar \psi}^+_u|\rho|\psi^-_u,{\bar \psi}^-_u,\psi^-_d,{\bar \psi}^-_d,\psi^-_s,{\bar \psi}^-_s,\Psi^-,{\bar \Psi}^-,Q^-,0>
\label{ncf}
\eea
where $Q_{\sigma \pm}^h(x)$ is the (quantum) gluon field and ${\overline \nabla}$ is defined by
\bea
{\bar \Psi}(x'){\overline \nabla }_{x'} \Psi(x')={\bar \Psi}(x')[{\vec \nabla }_{x'} \Psi(x')]-[{\vec \nabla}_{x'}{\bar \Psi}(x')] \Psi(x').
\label{nabl}
\eea
In eq. (\ref{ncf}) the $\rho$ is the initial density of state, $m^u,m^d,m^s,M$ are the mass of the up, down, strange, heavy quark, $|in>$ is the ground state in non-equilibrium QCD which is not the vacuum state $|0>$ and
\bea
F_{\sigma \lambda }^h[Q_\pm(x)]=\partial_\sigma Q_{\lambda \pm}^h(x) - \partial_\lambda Q_{\sigma \pm}^h(x)+gf^{hdb} Q_{\sigma \pm}^d(x) Q_{\lambda \pm}^b(x).
\label{fgx}
\eea
There are no ghost fields in this paper as we directly work with the ghost determinant ${\rm det}[ \frac{\delta \partial^\sigma Q_{\sigma +}^h}{\delta \omega^e_+}]$ in eq. (\ref{ncf}). Note that we work in the frozen ghost formalism at the initial time in non-equilibrium QCD \cite{kcn,kcn1}.

Although the proof of factorization of infrared divergence at fixed order in coupling constant is usually done in the diagrammatic method using pQCD, for example at LO, NLO, NNLO etc. in coupling constant, but it is useful to use the path integral formulation of QCD to prove the factorization of infrared divergence at all orders in coupling constant in QCD. Hence we will use the path integral formulation of non-equilibrium QCD developed in this section to prove NRQCD color octet factorization of P-wave heavy quarkonium production in non-equilibrium QCD at RHIC and LHC at all orders in coupling constant in this paper.

\section{ Infrared divergence and the light-like eikonal line in QCD}\label{infr}

In this section we describe the infrared divergence and the light-like eikonal line in QCD. We are interested in studying the factorization of infrared divergences at all orders in coupling constant for the process of a color octet heavy quark-antiquark pair in P-wave configuration interacting with the light-like eikonal line. The light-like eikonal line is characterized by the light-like velocity $l^\mu$ ($l^2=0$) where the the gauge link or the phase factor $L[x]$ associated with the light-like eikonal line is given by \cite{nkll}
\bea
L[x]={\cal P}e^{-igT^h \int_0^\infty ds~ l \cdot A^h(x+ls)}=e^{igT^h\omega^h(x)}
\label{llgl}
\eea
where $A_\sigma^h(x)$ is the SU(3) pure gauge field given by
\bea
T^hA_\sigma^h(x) =\frac{1}{ig}(\partial_\sigma L[x])L^{-1}[x].
\label{spgx}
\eea

In QED the Feynman diagram for the interaction of a photon with four-momentum $t^\mu$ interacting with the electron of four-momentum $w^\mu$ is given by
\bea
\frac{1}{{\not w}-m - {\not t}}{\not \epsilon}(t) u(w) =-\frac{w \cdot \epsilon(t)}{w\cdot t}u(w) + \frac{{\not t} {\not \epsilon}(t)}{2w\cdot t}u(w)={\cal M}_{eik}(t)+{\cal M}_{non-eik}(t)
\label{feik}
\eea
where ${\cal M}_{eik}$ is the eikonal part, ${\cal M}_{non-eik}$ is the non-eikonal part and $m$ is the mass of the electron. From eq. (\ref{feik}) we find in the infrared limit of the photon (in the limit $t^\mu \rightarrow 0$)
\bea
{\cal M}_{eik}^{\rm trans}(t)=0,~~~~~~~{\cal M}_{non-eik}^{\rm pure~gauge}(t)=0,~~~~~~~~~{\cal M}_{non-eik}^{\rm trans}(t)={\rm finite},~~~~~~~~{\cal M}_{eik}^{\rm pure~gauge}(t)\rightarrow \infty \nonumber \\
\label{feik1}
\eea
where the ${\cal M}_{eik}^{\rm trans}(t)$ means the eikonal part in eq. (\ref{feik}) is evaluated with the transverse polarization of the photon given by \cite{xyg}
\bea
\epsilon_\sigma^{\rm trans}(t)=\epsilon_\sigma(t)-k_\sigma \frac{w \cdot \epsilon(t)}{w\cdot t}u(w)
\label{trph}
\eea
and the ${\cal M}_{eik}^{\rm pure~gauge}(t)$ means the eikonal part in eq. (\ref{feik}) is evaluated with the longitudinal polarization (pure gauge part) of the photon given by \cite{xyg}
\bea
\epsilon_\sigma^{\rm pure~gauge}(t)=k_\sigma \frac{w \cdot \epsilon(t)}{w\cdot t}u(w),~~~~~~~~~\epsilon_\sigma(t)=\epsilon_\sigma^{\rm trans}(t)+\epsilon_\sigma^{\rm pure~gauge}(t).
\label{pgph}
\eea
From eq. (\ref{feik1}) we find that if the photon field is the pure gauge field then the infrared divergence in quantum field theory can be studied by using the eikonal approximation without modifying the finite part of the cross section. Now we will show that the light-like eikonal line produces pure gauge field in quantum field theory.

It is well known that in the classical electrodynamics the electric charge moving at the speed of light produces the U(1) pure gauge field at all the positions except at the positions perpendicular to the motion of the charge at the time of closest approach \cite{csg,npg1,npg2}. This result is also valid in quantum field theory which can be seen as follows.

By performing the path integration of the photon field in the presence of the light-like eikonal current density with light-like four-velocity $l^\mu$ ($l^2=0$) the effective lagrangian density ${\cal L}_{\rm effective}(x)$ is given by \cite{nksw}
\bea
{\cal L}^{\rm effective}(x)=0
\label{efph}
\eea
at all the positions except at the positions perpendicular to the motion of the charge at the time of closest approach. Similarly by performing the path integration of the photon field in the presence of the non-eikonal current density of four-momentum $q^\mu$ and eikonal current density of four velocity $l^\mu$ the effective (interaction) lagrangian density ${\cal L}_{\rm effective}^{\rm interaction}(x)$ is given by \cite{nksw}
\bea
{\cal L}^{\rm effective}_{\rm interaction}(x)=0
\label{intph}
\eea
at all the positions except at the positions perpendicular to the motion of the charges at the time of closest approach.

Hence we find from eqs. (\ref{efph}) and (\ref{intph}) that the light-like eikonal line produces pure gauge field at all the positions except at the positions perpendicular to the motion of the charge at the time of closest approach in quantum field theory which agrees with the corresponding result in the classical mechanics \cite{csg,npg1,npg2}.

From eqs. (\ref{feik1}), (\ref{efph}) and (\ref{intph}) we find that the study of the infrared divergences due to the interaction of the photons with the light-like eikonal line in QED is enormously simplified by using the U(1) pure gauge background field \cite{nksw}.

The eikonal Feynman rule in QCD is same as that in QED except for the presence of the $T^a$ matrices. Hence the study of the infrared divergences due to the interaction of the gluons with the light-like eikonal line in QCD is enormously simplified by using the SU(3) pure gauge background field \cite{nksw}. 

In QCD the SU(3) pure gauge background field $A_\sigma^h(x)$ is given by eq. (\ref{spgx}).

\section{ Infrared divergences in NRQCD color octet P-wave heavy quarkonium production}\label{infr1}

In this section we discuss the infrared (IR) divergences in NRQCD color octet P-wave heavy quarkonium production. First of all note that the ultra violet (UV) divergence of NRQCD is different from the ultra violet (UV) of QCD. This is because in NRQCD heavy quarkonium formulation the ultra violet (UV) cut-off is taken to be $\sim M$ where $M$ is the mass of the heavy quark.

However, the infrared behavior in NRQCD and the infrared behavior in QCD remains same. Hence the infrared (IR) divergences in NRQCD heavy quarkonium production can be studied by using the eikonal Feynman rules of QCD which we developed in the section \ref{infr}.

In \cite{nks,nksw} we have considered the color octet heavy quark-antiquark pair in the S-wave configuration in the presence of the light-like eikonal line of four-velocity $l^\mu$ in QCD in vacuum. We have proved the factorization of the infrared divergences for this process and have shown that the NRQCD color octet non-perturbative matrix element of the S-wave heavy quarkonium production in QCD in vacuum is independent of the light-like four-velocity $l^\mu$ of the eikonal line.

In \cite{nkpw} we have considered the color octet heavy quark-antiquark pair in the P-wave configuration in the presence of the light-like eikonal line of four-velocity $l^\mu$ in QCD in vacuum. We have proved the factorization of the infrared divergences for this process and have shown that the NRQCD color octet non-perturbative matrix element of the P-wave heavy quarkonium production in QCD in vacuum is independent of the light-like four-velocity $l^\mu$ of the eikonal line.

In this paper we consider the color octet heavy quark-antiquark pair in the P-wave configuration in the presence of the light-like eikonal line in non-equilibrium QCD. As shown in section \ref{infr} the study of the infrared (IR) divergences due to the gluons interaction with the light-like eikonal line in QCD can be enormously simplified by using the SU(3) pure gauge background field as given by eq. (\ref{spgx}). Hence we will use the path integral formulation of QCD in the presence of SU(3) pure gauge background field as given by eq. (\ref{spgx}) to prove NRQCD color octet factorization of P-wave heavy quarkonium production in non-equilibrium QCD at all orders in coupling constant. We will show that the NRQCD color octet non-perturbative matrix element of the P-wave heavy quarkonium production in non-equilibrium QCD is independent of the light-like four-velocity $l^\mu$ of the eikonal line.

\section{Proof of NRQCD color octet factorization of P-wave heavy quarkonium production in non-equilibrium QCD at all orders in coupling constant}

In this section we will prove NRQCD color octet factorization of P-wave heavy quarkonium production in non-equilibrium QCD at all orders in coupling constant. In sections \ref{infr} and \ref{infr1} we saw that the infrared (IR) divergences in NRQCD/QCD at all orders in coupling constant due to the presence of light-like eikonal line can be studied by using the path integral formulation of QCD in the presence of the SU(3) pure gauge background field.

Extending eq. (\ref{ncf}) we find that the non-perturbative correlation function of the heavy quark-antiquarks of the type $<in|{\bar \Psi}_{kr}(x'){\overline \nabla }_{x'}T^c_{kl} \Psi_{lr}(x') {\bar \Psi}_{js}(x''){\overline \nabla }_{x''}T^c_{jn} \Psi_{ns}(x'')|in>_A$ in non-equilibrium QCD in the presence of the background field $A_\sigma^h(x)$ is given by \cite{xbt,kcn,kcn1}
\bea
&&<in|{\bar \Psi}_{kr}(x'){\overline \nabla }_{x'}T^c_{kl} \Psi_{lr}(x') {\bar \Psi}_{js}(x''){\overline \nabla }_{x''}T^c_{jn} \Psi_{ns}(x'')|in>_A=\int [d{\bar \psi}^u_+] [d{\psi}^u_+][d{\bar \psi}^u_-] [d{\psi}^u_-][d{\bar \psi}^d_+] [d{\psi}^d_+]\nonumber \\
&&[d{\bar \psi}^d_-] [d{\psi}^d_-][d{\bar \psi}^s_+] [d{\psi}^s_+][d{\bar \psi}^s_-] [d{\psi}^s_-][d{\bar \Psi}_+] [d{\Psi}_+][d{\bar \Psi}_-] [d{\Psi}_-]\nonumber \\
&&\times {\bar \Psi}_{kr}(x'){\overline \nabla }_{x'}T^c_{kl} \Psi_{lr}(x') {\bar \Psi}_{js}(x''){\overline \nabla }_{x''}T^c_{jn} \Psi_{ns}(x'')\times {\rm det}[ \frac{\delta B_{ +}^h}{\delta \omega^e_+}]\times {\rm det}[ \frac{\delta B_{ -}^h}{\delta \omega^e_-}]\nonumber \\
&& \times ~{\rm exp}[i\int d^4x [-\frac{1}{4} F_{\sigma \lambda }^h[Q_+(x)+A_+(x)]F^{\sigma \lambda h}[Q_+(x)+A_+(x)] \nonumber \\
&&+\frac{1}{4} F_{\sigma \lambda }^h[Q_-(x)+A_-(x)]F^{\sigma \lambda h}[Q_-(x)+A_-(x)]    -\frac{1}{2\gamma} [B_{ +}^h(x)]^2\nonumber \\
&&+\frac{1}{2\gamma} [B_{ -}^h(x)]^2  +{\bar \psi}^u_{k+}(x)[\delta^{kl}(i{\not \partial} -m^u)+gT^h_{kl}({\not Q}_+^h(x)+{\not A}_+^h(x))]\psi^u_{l+}(x)\nonumber \\
&& -{\bar \psi}^u_{k-}(x)[\delta^{kl}(i{\not \partial} -m^u)+gT^h_{kl}({\not Q}_-^h(x)+{\not A}_-^h(x))]\psi^u_{l-}(x)  \nonumber \\
&& +{\bar \psi}^d_{k+}(x)[\delta^{kl}(i{\not \partial} -m^d)+gT^h_{kl}({\not Q}_+^h(x)+{\not A}_+^h(x))]\psi^d_{l+}(x)\nonumber \\
&& -{\bar \psi}^d_{k-}(x)[\delta^{kl}(i{\not \partial} -m^d)+gT^h_{kl}({\not Q}_-^h(x)+{\not A}_-^h(x))]\psi^d_{l-}(x) \nonumber \\
&&+{\bar \psi}^s_{k+}(x)[\delta^{kl}(i{\not \partial} -m^s)+gT^h_{kl}({\not Q}_+^h(x)+{\not A}_+^h(x))]\psi^s_{l+}(x)\nonumber \\
&& -{\bar \psi}^s_{k-}(x)[\delta^{kl}(i{\not \partial} -m^s)+gT^h_{kl}({\not Q}_-^h(x)+{\not A}_-^h(x))]\psi^s_{l-}(x)  \nonumber \\
&&+{\bar \Psi}_{k+}(x)[\delta^{kl}(i{\not \partial} -M)+gT^h_{kl}({\not Q}_+^h(x)+{\not A}_+^h(x))]\Psi_{l+}(x)\nonumber \\
&& -{\bar \Psi}_{k-}(x)[\delta^{kl}(i{\not \partial} -M)+gT^h_{kl}({\not Q}_-^h(x)+{\not A}_-^h(x))]\Psi_{l-}(x)  ]]\nonumber \\
&&\times <0,Q^++A^+,\Psi^+,{\bar \Psi}^+,\psi^+_s,{\bar \psi}^+_s,\psi^+_d,{\bar \psi}^+_d,\psi^+_u,{\bar \psi}^+_u|\rho|\psi^-_u,{\bar \psi}^-_u,\psi^-_d,{\bar \psi}^-_d,\psi^-_s,{\bar \psi}^-_s,\Psi^-,{\bar \Psi}^-,Q^-+A^-,0>\nonumber \\
\label{ancf}
\eea
where the background gauge fixing $B^h_\pm(x)$ is given by
\bea
B^h_\pm(x)=\partial^\sigma Q_{\sigma \pm}^h(x) +gf^{hdb} A_{\sigma \pm}^d(x)Q^{\sigma b}_\pm(x)
\label{xbf}
\eea
along with the (infinitesimal) type I gauge transformation \cite{xbt,xtf,xsz}
\bea
\delta A_{\sigma \pm}^h(x) = gf^{hdb} A_{\sigma \pm}^d(x) \omega^b_\pm(x)+\partial_\sigma \omega^h_\pm(x),~~~~~~~~~~~~~~\delta Q_{\sigma \pm}^h(x) = gf^{hdb} Q_{\sigma \pm}^d(x) \omega^b_\pm(x).
\label{xtp1}
\eea
The non-abelian field tensor $F_{\sigma \lambda }^h[Q_\pm(x)+A_\pm(x)]$ in the background field method of QCD in non-equilibrium in eq. (\ref{ancf}) is given by
\bea
&&F_{\sigma \lambda }^h[Q_\pm(x)+A_\pm(x)]\nonumber \\
&&=\partial_\sigma (Q_{\lambda \pm}^h(x) +A_{\lambda \pm}^h(x))- \partial_\lambda (Q_{\sigma \pm}^h(x)+A_{\sigma \pm}^h(x)))+gf^{hdb} (Q_{\sigma \pm}^d(x) +A_{\sigma \pm}^d(x))(Q_{\lambda \pm}^b(x)+A_{\lambda \pm}^b(x)).\nonumber \\
\label{afgx}
\eea
Note that there are no ghost fields in eq. (\ref{ancf}) as we directly work with the ghost determinant ${\rm det}[ \frac{\delta B_{ +}^h}{\delta \omega^e_+}]$. 

Changing the integration variable $Q_\pm(x)\rightarrow Q_\pm(x)-A_\pm(x)$ in eq. (\ref{ancf}) we find
\bea
&&<in|{\bar \Psi}_{kr}(x'){\overline \nabla }_{x'}T^c_{kl} \Psi_{lr}(x') {\bar \Psi}_{js}(x''){\overline \nabla }_{x''}T^c_{jn} \Psi_{ns}(x'')|in>_A=\int [d{\bar \psi}^u_+] [d{\psi}^u_+][d{\bar \psi}^u_-] [d{\psi}^u_-][d{\bar \psi}^d_+] [d{\psi}^d_+]\nonumber \\
&&[d{\bar \psi}^d_-] [d{\psi}^d_-][d{\bar \psi}^s_+] [d{\psi}^s_+][d{\bar \psi}^s_-] [d{\psi}^s_-][d{\bar \Psi}_+] [d{\Psi}_+][d{\bar \Psi}_-] [d{\Psi}_-]\nonumber \\
&&\times {\bar \Psi}_{kr}(x'){\overline \nabla }_{x'}T^c_{kl} \Psi_{lr}(x') {\bar \Psi}_{js}(x''){\overline \nabla }_{x''}T^c_{jn} \Psi_{ns}(x'')\times {\rm det}[ \frac{\delta B_{f +}^h}{\delta \omega^e_+}]\times {\rm det}[ \frac{\delta B_{f -}^h}{\delta \omega^e_-}]\nonumber \\
&& \times ~{\rm exp}[i\int d^4x [-\frac{1}{4} F_{\sigma \lambda }^h[Q_+(x)]F^{\sigma \lambda h}[Q_+(x)] +\frac{1}{4} F_{\sigma \lambda }^h[Q_-(x)]F^{\sigma \lambda h}[Q_-(x)]    -\frac{1}{2\gamma} [B_{f +}^h(x)]^2\nonumber \\
&&+\frac{1}{2\gamma} [B_{f -}^h(x)]^2  +{\bar \psi}^u_{k+}(x)[\delta^{kl}(i{\not \partial} -m^u)+gT^h_{kl}{\not Q}_+^h(x)]\psi^u_{l+}(x)\nonumber \\
&& -{\bar \psi}^u_{k-}(x)[\delta^{kl}(i{\not \partial} -m^u)+gT^h_{kl}{\not Q}_-^h(x)]\psi^u_{l-}(x)   +{\bar \psi}^d_{k+}(x)[\delta^{kl}(i{\not \partial} -m^d)+gT^h_{kl}{\not Q}_+^h(x)]\psi^d_{l+}(x)\nonumber \\
&& -{\bar \psi}^d_{k-}(x)[\delta^{kl}(i{\not \partial} -m^d)+gT^h_{kl}{\not Q}_-^h(x)]\psi^d_{l-}(x) +{\bar \psi}^s_{k+}(x)[\delta^{kl}(i{\not \partial} -m^s)+gT^h_{kl}{\not Q}_+^h(x)]\psi^s_{l+}(x)\nonumber \\
&& -{\bar \psi}^s_{k-}(x)[\delta^{kl}(i{\not \partial} -m^s)+gT^h_{kl}{\not Q}_-^h(x)]\psi^s_{l-}(x)  +{\bar \Psi}_{k+}(x)[\delta^{kl}(i{\not \partial} -M)+gT^h_{kl}{\not Q}_+^h(x)]\Psi_{l+}(x)\nonumber \\
&& -{\bar \Psi}_{k-}(x)[\delta^{kl}(i{\not \partial} -M)+gT^h_{kl}{\not Q}_-^h(x)]\Psi_{l-}(x)  ]]\nonumber \\
&&\times <0,Q^+,\Psi^+,{\bar \Psi}^+,\psi^+_s,{\bar \psi}^+_s,\psi^+_d,{\bar \psi}^+_d,\psi^+_u,{\bar \psi}^+_u|\rho|\psi^-_u,{\bar \psi}^-_u,\psi^-_d,{\bar \psi}^-_d,\psi^-_s,{\bar \psi}^-_s,\Psi^-,{\bar \Psi}^-,Q^-,0>
\label{bncf}
\eea
where
\bea
B^h_{f \pm}(x)=\partial^\sigma Q_{\sigma \pm}^h(x) +gf^{hdb} A_{\sigma \pm}^d(x)Q^{\sigma b}_\pm(x)-\partial^\sigma A_{\sigma \pm}^h(x)
\label{ybf}
\eea
and eq. (\ref{xtp1}) becomes
\bea
\delta Q_{\sigma \pm}^h(x) = gf^{hdb} Q_{\sigma \pm}^d(x) \omega^b_\pm(x)+\partial_\sigma \omega^h_\pm(x).
\label{xtp2}
\eea
The gauge transformation of the quark field is given by (see eq. (\ref{llgl}))
\bea
\psi'^k_{\pm}(x) =[e^{igT^h \omega^h_\pm(x)}]_{kl}\psi^l_\pm(x)=[{\cal P}e^{-igT^h \int_0^\infty ds~ l \cdot A^h_\pm(x+ls)}]_{kl}\psi^l_\pm(x)=[L_\pm(x)]_{kl}\psi^l_\pm(x).
\label{lmgl}
\eea
Since the primed and unprimed variables do not change the value of the integration we find from eq. (\ref{bncf}) that
\bea
&&<in|{\bar \Psi}_{kr}(x'){\overline \nabla }_{x'}T^c_{kl} \Psi_{lr}(x') {\bar \Psi}_{js}(x''){\overline \nabla }_{x''}T^c_{jn} \Psi_{ns}(x'')|in>_A=\int [d{\bar \psi}'^u_+] [d{\psi}'^u_+][d{\bar \psi}'^u_-] [d{\psi}'^u_-][d{\bar \psi}'^d_+] [d{\psi}'^d_+]\nonumber \\
&&[d{\bar \psi}'^d_-] [d{\psi}'^d_-][d{\bar \psi}'^s_+] [d{\psi}'^s_+][d{\bar \psi}'^s_-] [d{\psi}'^s_-][d{\bar \Psi}'_+] [d{\Psi}'_+][d{\bar \Psi}'_-] [d{\Psi}'_-]\nonumber \\
&&\times {\bar \Psi}'_{kr}(x'){\overline \nabla }_{x'}T^c_{kl} \Psi'_{lr}(x') {\bar \Psi}'_{js}(x''){\overline \nabla }_{x''}T^c_{jn} \Psi'_{ns}(x'')\times {\rm det}[ \frac{\delta B_{f +}'^h}{\delta \omega^e_+}]\times {\rm det}[ \frac{\delta B_{f -}'^h}{\delta \omega^e_-}]\nonumber \\
&& \times ~{\rm exp}[i\int d^4x [-\frac{1}{4} F_{\sigma \lambda }^h[Q'_+(x)]F^{\sigma \lambda h}[Q'_+(x)] +\frac{1}{4} F_{\sigma \lambda }^h[Q'_-(x)]F^{\sigma \lambda h}[Q'_-(x)]    -\frac{1}{2\gamma} [B_{f +}'^h(x)]^2\nonumber \\
&&+\frac{1}{2\gamma} [B_{f -}'^h(x)]^2  +{\bar \psi}'^u_{k+}(x)[\delta^{kl}(i{\not \partial} -m^u)+gT^h_{kl}{\not Q}_+'^h(x)]\psi'^u_{l+}(x)\nonumber \\
&& -{\bar \psi}'^u_{k-}(x)[\delta^{kl}(i{\not \partial} -m^u)+gT^h_{kl}{\not Q}_-'^h(x)]\psi'^u_{l-}(x)   +{\bar \psi}'^d_{k+}(x)[\delta^{kl}(i{\not \partial} -m^d)+gT^h_{kl}{\not Q}_+'^h(x)]\psi'^d_{l+}(x)\nonumber \\
&& -{\bar \psi}'^d_{k-}(x)[\delta^{kl}(i{\not \partial} -m^d)+gT^h_{kl}{\not Q}_-'^h(x)]\psi'^d_{l-}(x) +{\bar \psi}'^s_{k+}(x)[\delta^{kl}(i{\not \partial} -m^s)+gT^h_{kl}{\not Q}_+'^h(x)]\psi'^s_{l+}(x)\nonumber \\
&& -{\bar \psi}'^s_{k-}(x)[\delta^{kl}(i{\not \partial} -m^s)+gT^h_{kl}{\not Q}_-'^h(x)]\psi'^s_{l-}(x)  +{\bar \Psi}'_{k+}(x)[\delta^{kl}(i{\not \partial} -M)+gT^h_{kl}{\not Q}_+'^h(x)]\Psi'_{l+}(x)\nonumber \\
&& -{\bar \Psi}'_{k-}(x)[\delta^{kl}(i{\not \partial} -M)+gT^h_{kl}{\not Q}_-'^h(x)]\Psi'_{l-}(x)  ]]\nonumber \\
&&\times <0,Q'^+,\Psi'^+,{\bar \Psi}'^+,\psi'^+_s,{\bar \psi}'^+_s,\psi'^+_d,{\bar \psi}'^+_d,\psi'^+_u,{\bar \psi}'^+_u|\rho|\psi'^-_u,{\bar \psi}'^-_u,\psi'^-_d,{\bar \psi}'^-_d,\psi'^-_s,{\bar \psi}'^-_s,\Psi'^-,{\bar \Psi}'^-,Q'^-,0>.\nonumber \\
\label{cncf}
\eea
Since we work in the frozen ghost formalism at the initial time \cite{kcn,kcn1} the initial density of state $<0,Q^+,\Psi^+,{\bar \Psi}^+,\psi^+_s,{\bar \psi}^+_s,\psi^+_d,{\bar \psi}^+_d,\psi^+_u,{\bar \psi}^+_u|\rho|\psi^-_u,{\bar \psi}^-_u,\psi^-_d,{\bar \psi}^-_d,\psi^-_s,{\bar \psi}^-_s,\Psi^-,{\bar \Psi}^-,Q^-,0>$ is gauge invariant by definition, {\it i. e.},
\bea
&&<0,Q'^+,\Psi'^+,{\bar \Psi}'^+,\psi'^+_s,{\bar \psi}'^+_s,\psi'^+_d,{\bar \psi}'^+_d,\psi'^+_u,{\bar \psi}'^+_u|\rho|\psi'^-_u,{\bar \psi}'^-_u,\psi'^-_d,{\bar \psi}'^-_d,\psi'^-_s,{\bar \psi}'^-_s,\Psi'^-,{\bar \Psi}'^-,Q'^-,0>\nonumber \\
&&=<0,Q^+,\Psi^+,{\bar \Psi}^+,\psi^+_s,{\bar \psi}^+_s,\psi^+_d,{\bar \psi}^+_d,\psi^+_u,{\bar \psi}^+_u|\rho|\psi^-_u,{\bar \psi}^-_u,\psi^-_d,{\bar \psi}^-_d,\psi^-_s,{\bar \psi}^-_s,\Psi^-,{\bar \Psi}^-,Q^-,0>.
\label{xgv}
\eea
Using eq. (\ref{xgv}) in (\ref{cncf}) we find
\bea
&&<in|{\bar \Psi}_{kr}(x'){\overline \nabla }_{x'}T^c_{kl} \Psi_{lr}(x') {\bar \Psi}_{js}(x''){\overline \nabla }_{x''}T^c_{jn} \Psi_{ns}(x'')|in>_A=\int [d{\bar \psi}'^u_+] [d{\psi}'^u_+][d{\bar \psi}'^u_-] [d{\psi}'^u_-][d{\bar \psi}'^d_+] [d{\psi}'^d_+]\nonumber \\
&&[d{\bar \psi}'^d_-] [d{\psi}'^d_-][d{\bar \psi}'^s_+] [d{\psi}'^s_+][d{\bar \psi}'^s_-] [d{\psi}'^s_-][d{\bar \Psi}'_+] [d{\Psi}'_+][d{\bar \Psi}'_-] [d{\Psi}'_-]\nonumber \\
&&\times {\bar \Psi}'_{kr}(x'){\overline \nabla }_{x'}T^c_{kl} \Psi'_{lr}(x') {\bar \Psi}'_{js}(x''){\overline \nabla }_{x''}T^c_{jn} \Psi'_{ns}(x'')\times {\rm det}[ \frac{\delta B_{f +}'^h}{\delta \omega^e_+}]\times {\rm det}[ \frac{\delta B_{f -}'^h}{\delta \omega^e_-}]\nonumber \\
&& \times ~{\rm exp}[i\int d^4x [-\frac{1}{4} F_{\sigma \lambda }^h[Q'_+(x)]F^{\sigma \lambda h}[Q'_+(x)] +\frac{1}{4} F_{\sigma \lambda }^h[Q'_-(x)]F^{\sigma \lambda h}[Q'_-(x)]    -\frac{1}{2\gamma} [B_{f +}'^h(x)]^2\nonumber \\
&&+\frac{1}{2\gamma} [B_{f -}'^h(x)]^2  +{\bar \psi}'^u_{k+}(x)[\delta^{kl}(i{\not \partial} -m^u)+gT^h_{kl}{\not Q}_+'^h(x)]\psi'^u_{l+}(x)\nonumber \\
&& -{\bar \psi}'^u_{k-}(x)[\delta^{kl}(i{\not \partial} -m^u)+gT^h_{kl}{\not Q}_-'^h(x)]\psi'^u_{l-}(x)   +{\bar \psi}'^d_{k+}(x)[\delta^{kl}(i{\not \partial} -m^d)+gT^h_{kl}{\not Q}_+'^h(x)]\psi'^d_{l+}(x)\nonumber \\
&& -{\bar \psi}'^d_{k-}(x)[\delta^{kl}(i{\not \partial} -m^d)+gT^h_{kl}{\not Q}_-'^h(x)]\psi'^d_{l-}(x) +{\bar \psi}'^s_{k+}(x)[\delta^{kl}(i{\not \partial} -m^s)+gT^h_{kl}{\not Q}_+'^h(x)]\psi'^s_{l+}(x)\nonumber \\
&& -{\bar \psi}'^s_{k-}(x)[\delta^{kl}(i{\not \partial} -m^s)+gT^h_{kl}{\not Q}_-'^h(x)]\psi'^s_{l-}(x)  +{\bar \Psi}'_{k+}(x)[\delta^{kl}(i{\not \partial} -M)+gT^h_{kl}{\not Q}_+'^h(x)]\Psi'_{l+}(x)\nonumber \\
&& -{\bar \Psi}'_{k-}(x)[\delta^{kl}(i{\not \partial} -M)+gT^h_{kl}{\not Q}_-'^h(x)]\Psi'_{l-}(x)  ]]\nonumber \\
&&\times <0,Q^+,\Psi^+,{\bar \Psi}^+,\psi^+_s,{\bar \psi}^+_s,\psi^+_d,{\bar \psi}^+_d,\psi^+_u,{\bar \psi}^+_u|\rho|\psi^-_u,{\bar \psi}^-_u,\psi^-_d,{\bar \psi}^-_d,\psi^-_s,{\bar \psi}^-_s,\Psi^-,{\bar \Psi}^-,Q^-,0>
\label{dncf}
\eea
When the background field $A_{\sigma \pm}^h(x)$ is the SU(3) pure gauge background field as given by eq. (\ref{spgx}) we find from eqs. (\ref{xtp2}) and (\ref{lmgl}) \cite{nksw}
\bea
&&[d{\bar \psi}'^u_+] [d{\psi}'^u_+][d{\bar \psi}'^u_-] [d{\psi}'^u_-][d{\bar \psi}'^d_+] [d{\psi}'^d_+][d{\bar \psi}'^d_-] [d{\psi}'^d_-][d{\bar \psi}'^s_+] [d{\psi}'^s_+][d{\bar \psi}'^s_-] [d{\psi}'^s_-][d{\bar \Psi}'_+] [d{\Psi}'_+][d{\bar \Psi}'_-] [d{\Psi}'_-]\nonumber \\
&&\times {\rm det}[ \frac{\delta B_{f +}'^h}{\delta \omega^e_+}]\times {\rm det}[ \frac{\delta B_{f -}'^h}{\delta \omega^e_-}]\nonumber \\
&& \times ~{\rm exp}[i\int d^4x [-\frac{1}{4} F_{\sigma \lambda }^h[Q'_+(x)]F^{\sigma \lambda h}[Q'_+(x)] +\frac{1}{4} F_{\sigma \lambda }^h[Q'_-(x)]F^{\sigma \lambda h}[Q'_-(x)]    -\frac{1}{2\gamma} [B_{f +}'^h(x)]^2\nonumber \\
&&+\frac{1}{2\gamma} [B_{f -}'^h(x)]^2  +{\bar \psi}'^u_{k+}(x)[\delta^{kl}(i{\not \partial} -m^u)+gT^h_{kl}{\not Q}_+'^h(x)]\psi'^u_{l+}(x)\nonumber \\
&& -{\bar \psi}'^u_{k-}(x)[\delta^{kl}(i{\not \partial} -m^u)+gT^h_{kl}{\not Q}_-'^h(x)]\psi'^u_{l-}(x)   +{\bar \psi}'^d_{k+}(x)[\delta^{kl}(i{\not \partial} -m^d)+gT^h_{kl}{\not Q}_+'^h(x)]\psi'^d_{l+}(x)\nonumber \\
&& -{\bar \psi}'^d_{k-}(x)[\delta^{kl}(i{\not \partial} -m^d)+gT^h_{kl}{\not Q}_-'^h(x)]\psi'^d_{l-}(x) +{\bar \psi}'^s_{k+}(x)[\delta^{kl}(i{\not \partial} -m^s)+gT^h_{kl}{\not Q}_+'^h(x)]\psi'^s_{l+}(x)\nonumber \\
&& -{\bar \psi}'^s_{k-}(x)[\delta^{kl}(i{\not \partial} -m^s)+gT^h_{kl}{\not Q}_-'^h(x)]\psi'^s_{l-}(x)  +{\bar \Psi}'_{k+}(x)[\delta^{kl}(i{\not \partial} -M)+gT^h_{kl}{\not Q}_+'^h(x)]\Psi'_{l+}(x)\nonumber \\
&& -{\bar \Psi}'_{k-}(x)[\delta^{kl}(i{\not \partial} -M)+gT^h_{kl}{\not Q}_-'^h(x)]\Psi'_{l-}(x)  ]]\nonumber \\
&& =d{\bar \psi}^u_+] [d{\psi}^u_+][d{\bar \psi}^u_-] [d{\psi}^u_-][d{\bar \psi}^d_+] [d{\psi}^d_+][d{\bar \psi}^d_-] [d{\psi}^d_-][d{\bar \psi}^s_+] [d{\psi}^s_+][d{\bar \psi}^s_-] [d{\psi}^s_-][d{\bar \Psi}_+] [d{\Psi}_+][d{\bar \Psi}_-] [d{\Psi}_-]\nonumber \\
&&\times {\rm det}[ \frac{\delta \partial^\sigma Q_{\sigma +}^h}{\delta \omega^e_+}]\times {\rm det}[ \frac{\delta \partial^\sigma Q_{\sigma -}^h}{\delta \omega^e_-}]\nonumber \\
&& \times ~{\rm exp}[i\int d^4x [-\frac{1}{4} F_{\sigma \lambda }^h[Q_+(x)]F^{\sigma \lambda h}[Q_+(x)] +\frac{1}{4} F_{\sigma \lambda }^h[Q_-(x)]F^{\sigma \lambda h}[Q_-(x)]    -\frac{1}{2\gamma} [\partial^\sigma Q_{\sigma +}^h(x)]^2\nonumber \\
&&+\frac{1}{2\gamma} [\partial^\sigma Q_{\sigma -}^h(x)]^2  +{\bar \psi}^u_{k+}(x)[\delta^{kl}(i{\not \partial} -m^u)+gT^h_{kl}{\not Q}_+^h(x)]\psi^u_{l+}(x)\nonumber \\
&& -{\bar \psi}^u_{k-}(x)[\delta^{kl}(i{\not \partial} -m^u)+gT^h_{kl}{\not Q}_-^h(x)]\psi^u_{l-}(x)   +{\bar \psi}^d_{k+}(x)[\delta^{kl}(i{\not \partial} -m^d)+gT^h_{kl}{\not Q}_+^h(x)]\psi^d_{l+}(x)\nonumber \\
&& -{\bar \psi}^d_{k-}(x)[\delta^{kl}(i{\not \partial} -m^d)+gT^h_{kl}{\not Q}_-^h(x)]\psi^d_{l-}(x) +{\bar \psi}^s_{k+}(x)[\delta^{kl}(i{\not \partial} -m^s)+gT^h_{kl}{\not Q}_+^h(x)]\psi^s_{l+}(x)\nonumber \\
&& -{\bar \psi}^s_{k-}(x)[\delta^{kl}(i{\not \partial} -m^s)+gT^h_{kl}{\not Q}_-^h(x)]\psi^s_{l-}(x)  +{\bar \Psi}_{k+}(x)[\delta^{kl}(i{\not \partial} -M)+gT^h_{kl}{\not Q}_+^h(x)]\Psi_{l+}(x)\nonumber \\
&& -{\bar \Psi}_{k-}(x)[\delta^{kl}(i{\not \partial} -M)+gT^h_{kl}{\not Q}_-^h(x)]\Psi_{l-}(x)  ]].
\label{encf}
\eea
Using eqs. (\ref{xtp2}) and (\ref{encf}) in (\ref{dncf}) we find
\bea
&&<in|{\bar \Psi}_{kr}(x')L_r[x']{\overline \nabla }_{x'}T^c_{kl} L^{-1}_r[x']\Psi_{lr}(x') {\bar \Psi}_{js}(x'')L_r[x'']{\overline \nabla }_{x''}T^c_{jn} L^{-1}[x'']\Psi_{ns}(x'')|in>_A\nonumber \\
&&=\int [d{\bar \psi}^u_+] [d{\psi}^u_+][d{\bar \psi}^u_-] [d{\psi}^u_-][d{\bar \psi}^d_+] [d{\psi}^d_+][d{\bar \psi}^d_-] [d{\psi}^d_-][d{\bar \psi}^s_+] [d{\psi}^s_+][d{\bar \psi}^s_-] [d{\psi}^s_-][d{\bar \Psi}_+] [d{\Psi}_+][d{\bar \Psi}_-] [d{\Psi}_-]\nonumber \\
&&\times {\bar \Psi}_{kr}(x'){\overline \nabla }_{x'}T^c_{kl} \Psi_{lr}(x') {\bar \Psi}_{js}(x''){\overline \nabla }_{x''}T^c_{jn} \Psi_{ns}(x'')\times {\rm det}[ \frac{\delta \partial^\sigma Q_{\sigma +}^h}{\delta \omega^e_+}]\times {\rm det}[ \frac{\delta \partial^\sigma Q_{\sigma -}^h}{\delta \omega^e_-}]\nonumber \\
&& \times ~{\rm exp}[i\int d^4x [-\frac{1}{4} F_{\sigma \lambda }^h[Q_+(x)]F^{\sigma \lambda h}[Q_+(x)] +\frac{1}{4} F_{\sigma \lambda }^h[Q_-(x)]F^{\sigma \lambda h}[Q_-(x)]    -\frac{1}{2\gamma} [\partial^\sigma Q_{\sigma +}^h(x)]^2\nonumber \\
&&+\frac{1}{2\gamma} [\partial^\sigma Q_{\sigma -}^h(x)]^2  +{\bar \psi}^u_{k+}(x)[\delta^{kl}(i{\not \partial} -m^u)+gT^h_{kl}{\not Q}_+^h(x)]\psi^u_{l+}(x)\nonumber \\
&& -{\bar \psi}^u_{k-}(x)[\delta^{kl}(i{\not \partial} -m^u)+gT^h_{kl}{\not Q}_-^h(x)]\psi^u_{l-}(x)   +{\bar \psi}^d_{k+}(x)[\delta^{kl}(i{\not \partial} -m^d)+gT^h_{kl}{\not Q}_+^h(x)]\psi^d_{l+}(x)\nonumber \\
&& -{\bar \psi}^d_{k-}(x)[\delta^{kl}(i{\not \partial} -m^d)+gT^h_{kl}{\not Q}_-^h(x)]\psi^d_{l-}(x) +{\bar \psi}^s_{k+}(x)[\delta^{kl}(i{\not \partial} -m^s)+gT^h_{kl}{\not Q}_+^h(x)]\psi^s_{l+}(x)\nonumber \\
&& -{\bar \psi}^s_{k-}(x)[\delta^{kl}(i{\not \partial} -m^s)+gT^h_{kl}{\not Q}_-^h(x)]\psi^s_{l-}(x)  +{\bar \Psi}_{k+}(x)[\delta^{kl}(i{\not \partial} -M)+gT^h_{kl}{\not Q}_+^h(x)]\Psi_{l+}(x)\nonumber \\
&& -{\bar \Psi}_{k-}(x)[\delta^{kl}(i{\not \partial} -M)+gT^h_{kl}{\not Q}_-^h(x)]\Psi_{l-}(x)  ]]\nonumber \\
&&\times <0,Q^+,\Psi^+,{\bar \Psi}^+,\psi^+_s,{\bar \psi}^+_s,\psi^+_d,{\bar \psi}^+_d,\psi^+_u,{\bar \psi}^+_u|\rho|\psi^-_u,{\bar \psi}^-_u,\psi^-_d,{\bar \psi}^-_d,\psi^-_s,{\bar \psi}^-_s,\Psi^-,{\bar \Psi}^-,Q^-,0>
\label{fncf}
\eea
which gives by using eq. (\ref{ncf})
\bea
&&<in|{\bar \Psi}_{kr}(x')L_r[x']{\overline \nabla }_{x'}T^c_{kl} L^{-1}_r[x']\Psi_{lr}(x') {\bar \Psi}_{js}(x'')L_r[x'']{\overline \nabla }_{x''}T^c_{jn} L^{-1}[x'']\Psi_{ns}(x'')|in>_A\nonumber \\
&&=<in|{\bar \Psi}_{kr}(x'){\overline \nabla }_{x'}T^c_{kl} \Psi_{lr}(x') {\bar \Psi}_{js}(x''){\overline \nabla }_{x''}T^c_{jn} \Psi_{ns}(x'')|in>
\label{gncf}
\eea
which proves NRQCD color octet factorization of P-wave heavy quarkonium production in non-equilibrium QCD at all orders in coupling constant where the gauge link $L[x]$ in the fundamental representation of SU(3) is given by eq. (\ref{llgl}). Note that the repeated closed-time path indices $r,s$ are not summed in eq. (\ref{gncf}).

\section{Definition of NRQCD color octet non-perturbative matrix element of the P-wave heavy quarkonium production in non-equilibrium QCD}

From eq. (\ref{gncf}) we find that the definition of the NRQCD color octet non-perturbative matrix element of the P-wave heavy quarkonium production in non-equilibrium QCD at RHIC and LHC which is gauge invariant and is consistent with the factorization of the infrared (IR) divergences at all orders in coupling constant is given by
\bea
<in|{\cal O}_H|in> = <in|\xi^\dagger[0] L[0]{\overline \nabla }T^h L^{-1}[0]\Lambda[0] a^\dagger_H a_H \Lambda^\dagger[0] L[0] T^h {\overline \nabla } L^{-1}[0] \xi[0]|in>
\label{npm}
\eea
where $\Lambda~(\xi)$ is the two component Dirac spinor field that annihilates (creates) a heavy quark, $a_H$ is the annihilation operator of the heavy quarkonium, the gauge link $L[x]$ is given by eq. (\ref{llgl}) and the ${\overline \nabla }$ is defined in eq. (\ref{nabl}).

Since the right hand side of the eq. (\ref{gncf}) is independent of $l^\mu$ we find that the definition of the NRQCD color octet non-perturbative matrix element of the P-wave heavy quarkonium production in non-equilibrium QCD at RHIC and LHC in eq. (\ref{npm}) is independent of the light-like four-vector $l^\mu$ that appears in the gauge link $L[x]$ in eq. (\ref{llgl}).
\section{Conclusions}
Recently we have proved NRQCD color octet factorization of S-wave and P-wave heavy quarkonia production at all orders in coupling constant in QCD in vacuum at high energy colliders in \cite{nksw} and in \cite{nkpw} respectively. In this paper we have extended this to prove NRQCD color octet factorization of P-wave heavy quarkonium production in non-equilibrium QCD at RHIC and LHC at all orders in coupling constant. This proof is necessary to study the quark-gluon plasma at RHIC and LHC \cite{cn1,cn2,cn3,cn4}.

\end{document}